\documentclass[aps,preprint]{revtex4}
\usepackage{graphicx}%
\usepackage{dcolumn}
\usepackage{amsmath}

\newcommand{\lessim} {\mathop{\,<\kern - 1.05 em \lower 1.ex \hbox {$\sim$}\,}}
\newcommand{\grtsim} {\mathop {\,> \kern - 1.05 em \lower 1.ex \hbox 
{$\sim$}\,}}

\begin{document}
\normalsize  \title{Angle--Resolved Loss of Landau 
Quasiparticles in 2D Hubbard Model}

\author{Dra\v{z}en Zanchi}
\affiliation{Laboratoire de Physique Th\'eorique et Hautes Energies. 
Universit\'es Paris VI Pierre et Marie Currie -- Paris VII Denis Diderot, 
2 Place Jussieu, 75252 Paris C\'edex 05, France.}

\widetext

\begin{abstract}
The problem of weakly correlated electrons on a square lattice is
studied theoretically.  
A simple renormalization group scheme for the angle--resolved 
weight $Z(\theta)$ of the quasiparticles at the Fermi surface is presented
and applied to the Hubbard model.
Upon reduction of the  cutoff the Fermi surface is 
progressively destroyed from the van Hove points toward
the zone diagonals. 
Due to the renormalized $Z(\theta)$, divergences of both 
antiferromagnetic and superconducting correlation
functions are suppressed at the critical scale, where the
interactions  diverge.

\end{abstract}
\bigskip
LPTHE/01-17 \hspace{106mm}April 2001
\bigskip

\maketitle

\bigskip



\newpage


Understanding of the one--particle spectrum of strongly correlated systems near the 
metal--insulator transition is an extremely difficult task, particularly if one 
wants to construct a microscopic theory. One standard example is the pseudogap 
regime of the HTC superconductors. 
ARPES measurements \cite{ARPES_psgap}  showed that the Fermi surface 
is destroyed by 
correlations. This happens first near the 
van Hove points where the one-particle spectrum  develops the
characteristic 2-peak structure. Remaining parts 
of the Fermi surface, often called Fermi patches,  
get progressively narrower around Brillouin zone diagonals 
as the temperature decreases. 
Regions around van Hove points  contain
non--Fermi liquid with a pseudogap and other signatures of strong correlations  
such as flat bands.\cite{flat_band} In other words, the pseudogap has the
form similar to the absolute value of the $d_{x^2-y^2}$--superconducting (SC)
order--parameter. This vision of the pseudogap is in agreement with STM
results \cite{tunel_psgap} as well, regardless of details on how these
results are interpreted.
The above experiments still do not reveal much on the
origins of the pseudogap, namely whether it is simply the signature
of a liquid of pre--formed pairs
or something much richer in fluctuations. In fact it is known that 
the antiferromagnetic (AF) fluctuations are  also strong in the pseudogap 
regime.\cite{neutrons}

If we assume that AF and SC fluctuations are somehow {\em together} the
major
reason for the strong renormalization of the one--particle selfenergy, 
and for the consequent
partial destruction of the Fermi surface, then a many--body analysis
of the one--particle propagator can be done in a controlled way. In fact the 
weak coupling theory easily reproduces AF and SC fluctuations from particle--hole
(p-h) and particle--particle (p-p) loop--logarithms. Even if the coupling in 
realistic HTC systems is of the order of Fermi energy
(i.e. intermediate--to--strong), already the weak--coupling theory contains
the observed two--particle correlations. 
In this paper I answer the question of how the angle--resolved quasiparticle
weight $Z(\theta)$ is renormalized by strong and coupled AF and SC
fluctuations, and of the main consequences of the renormalized $Z(\theta)$
 to the characteristic angle--resolved two--particle 
correlation functions. The 
renormalization of the
quasiparticle weight in 2D was recently
 studied by Kishine and Yonemitsu.\cite{Kishine99}
To calculate the renormalization of
$Z$ resolved in the 
position on a flat Fermi surface
they used a two--loop selfenergy expansion with the two--loop--renormalized
vertex. 
The results show clearly that the flatness of the Fermi surface induces
the suppression of 
the quasiparticle residue and that this 
effect is anisotropic.
In the present work I consider the whole square 
Fermi surface of the 
Hubbard model. For this purpose I employ the
N--patch renormalization group theory.

After recent theoretical studies from several groups 
it emerged that the N--patch model describes in a 
systematic and
controlled way  weakly correlated electrons
near half--filling, and explains the  major aspects of
HTC-s.\cite{these,doucot_97,ZS_prb_00,Halboth_00,Honerkamp_01,Tsai_Marston}
Until now the RG analysis of the N--patch model has been done only on the
level of the two--particle scattering amplitudes or, in field--theoretical
jargon, of the four--point vertex $U(K_1,K_2,K_3)$. The analysis of the
renormalization group flow of $U$ as a
function of three patch indices $i_1,i_2,i_3$ gave 
several important results. 
Typically the amplitudes $U(i_1,i_2,i_3)$ diverge at some interaction-- and 
doping--dependent critical energy scale $\Lambda_c$.
For the case of the Hubbard model, we distinguish two main renormalization
regimes \cite{ZS_prb_00}, the {\em parquet regime} and the {\em BCS regime}.

In the  parquet regime ($|\mu| < \Lambda$)  both particle-particle and particle-hole
propagators have strong
contributions to the beta--function due to the van Hove singularities and
the Fermi surface nesting.  In this regime, 
provided $\Lambda \rightarrow \Lambda_c$
and neglecting the selfenrgy corrections,
 both SC and AF tendencies 
are strong and build {\em divergent}
correlation functions $\chi ^{SC}$ and $\chi ^{AF}$.
 The dominant component of the antiferromagnetic  
susceptibility is of the $s$--type and the dominant component of the
superconducting one is of the $d_{x^2-y^2}$--type. 
Both static compressibility $\chi _c$
and homogeneous magnetic susceptibility $\chi _s$ go to zero as the cutoff
approaches to 
its critical value $\Lambda_c$.\cite{these,Halboth_00,Honerkamp_01}
Consequently $\Lambda_c$ is the energy scale of
the crossover  between
the strange metal and the strongly correlated regime with gap or pseudogap.
A question arising from already existing results on the Hubbard
 model \cite{these,ZS_prb_00} and on its
extensions \cite{Halboth_00,Honerkamp_01,Tsai_Marston} is the following:
if one wants to
interpret the critical scale $\Lambda_c$ (or temperature) 
as the energy (temperature) $T^*$ 
for the onset of the pseudogap, why then are all signatures of the pseudogap 
not seen? This means in particular that 
 $\chi ^{AF}$ and $\chi ^{SC}$ should be finite and not diverging.
Emergent is the necessity to calculate the correlation functions with the
corrections due to the one--particle selfenergy.

At stronger doping ($|\mu| > \Lambda$) 
nesting properties get weaker so
that eventually only remaining renormalization channel is superconducting
(p-p). This is the BCS regime. 
There the superconductivity is simply BCS--like
with  the coupling constants and the angular profile of the 
order parameter  determined at
higher scales by a parquet--like flow, where $\Lambda$ was larger than 
the chemical potential.

Equivalent to the RG approach is the fast parquet theory
\cite{DzYak,Zheleznyak}, where $\theta$ variable (continuum version of
the patch index) is called the fast variable in addition to  the cutoff
logarithm called the slow variable. In two dimensions 
the parquet integro--differential equations always have 
mobile pole solutions, i.e. the AF and SC
fluctuations  decouple one from the other. This seems  to be 
in disagreement with the RG results,
where we detected  only immobile poles, the type of solution in which
all scattering amplitudes develop the pole at the same scale $\Lambda_c$.
The question of the consilience between the two theories is still
controversial.
However, the results of De Abreu and Dou\c{c}ot \cite{doucot_00} indicate 
that the mobile pole
solution is dominant only in the very vicinity of $\Lambda_c$ and that 
the fixed pole
solution is an {\em intermediate} solution, valid over several decades of
energy scale. The width of the ``very vicinity'' characterized by the mobile
poles depends on the coupling 
constant so that for reasonable and not too weak $U_0$ the final regime is so
close to $\Lambda_c$, that the couplings are already too strong and out of
reach of a  weak coupling theory. Consequently the real physical
interpretation can be given only to the immobile pole regime.
We will  concentrate on this  ``intermediate''
regime in which all fluctuations are coupled and at least behave 
as if having an immobile pole.

We suppose that the electronic Green function has the form
\begin{equation} \label{GF_Z}
G_l(K)=\frac{Z_l(\theta)}{i\omega-\xi({\bf k})}\; .
\end{equation}
$Z_l(\theta)$ is the angle--resolved scale dependent 
quasiparticle weight and $\xi({\bf k})$ is
the tight--binding dispersion. The formalism keeps the notation introduced
in the reference \cite{ZS_prb_00}. The form (\ref{GF_Z}) contains two main
approximations. The first one is to keep trace only of the renormalization of the
coherent part of the propagator. The second approximation is to assume that the
spectrum $\xi({\bf k})$ remains non--renormalized. This assumption implies
that we ignore the flow of the Fermi surface (FS) and of the Fermi
velocity. Because of the particle--hole symmetry the
flow of the FS is zero at  half--filling, where we can expect that
our form of the Green function is closer to reality than in the imperfectly
nested (non--half filled) case.

The flow equation for $Z(\theta)$ is derived from  the general and exact 
one--loop RG equation for the complete selfenergy $\Sigma(K)$ given in 
\cite{ZS_prb_00}. Let us suppose that we are at some scale $l$, and that we
know the propagator (\ref{GF_Z}). We integrate $dl$ further and look what
is the effective two--point vertex $\Gamma _2$ in the effective action
$S(l+dl)$; it is
\begin{equation} \label{Gamma_2}
\Gamma_2(l+dl)=Z_l^{-1}(\theta)(i\omega-\xi({\bf k}))+d\Sigma_l(K)\; .
\end{equation}
To find $Z_{l+dl}$ we expand $d\Sigma_l(K)$ in first order in $i\omega$ to
obtain
\begin{equation} \label{noviZ}
Z_{l+dl}=Z_l(1-Z_l\partial_{i\omega}d\Sigma) \; .
\end{equation}
The differential equation for $Z$ follows immediately
\begin{equation} \label{DlZ}
\partial _lZ_l(\theta)=-Z_l^2(\theta)\, \partial _{i\omega}\left[ \partial _l
\Sigma(K)\right] \arrowvert _{\xi=i\omega=0  }     \; .
\end{equation}
Only the terms of $\partial_l\Sigma$ which are linear in energy  contribute. 
These are just the terms which are marginal upon zeroth order scaling in
Shankar's sense.\cite{Shankar} We will  look for these terms. 

The equation for
$\partial _l\Sigma$ can be written as 
\begin{equation} \label{DlS}
\partial _l\Sigma(\theta,\epsilon,\omega)=\frac{\Lambda}{(2\pi)^2}\int
\frac{d\omega'}{2\pi}\sum_{\nu}\int{\cal J}_\nu(\theta',\Lambda)d\theta'\;
G_l(\theta',\omega',\nu\Lambda){\cal D}_l(K,K'_{\nu}) \; ,
\end{equation}
where ${\cal D}_l=2F_l-\tilde{F}_l$; $F_l$ and $\tilde{F}_l$ are 
energy--momenta dependent forward
and  backward scattering processes at the scale $l$,
related to the effective interaction in a way that $F_l(K_1,K_2)=U(K_1,K_2,K_1)$ and
$\tilde{F}_l(K_1,K_2)=U(K_1,K_2,K_2)$. ${\cal J}_\nu(\theta,\Lambda)$ is
the angle--resolved density of states at  
energy $\xi=\nu\Lambda$.
There is another, approximate but physically justified way to 
decompose ${\cal D}$. In fact, we will {\em suppose} that $D_l$ can be
written as a sum of p-p and p-h terms:
\begin{equation} \label{Decomp}
{\cal D}_l(K,K')={\cal D}_l^{pp}(K+K')+{\cal D}_l^{ph}(K-K')\; .
\end{equation}
The p-p part of the propagator ${\cal D}$ depends only on the total
energy--momentum 
$Q_{pp}=(\omega _{pp},{\bf q}_{pp})\equiv K+K'$ while the p-h part depends only on the 
energy--momentum transfer $Q_{ph}=(\omega _{ph},
{\bf q}_{ph})\equiv K-K'$. As usual we skip the marginal
part of the dependence on ${\bf q}_{pp}$ and ${\bf q}_{ph}$. 
For that purpose we note that both momenta 
can be written in the form 
\begin{equation} \label{both_q}
{\bf q}={\bf q}^{(0)}(\theta,\theta')+{\bf q}^{(1)}(\theta,\theta',\xi,\xi')\; ,
\end{equation}
where ${\bf q}$ stands either for ${\bf q}_{pp}$ or for ${\bf q}_{ph}$,
${\bf q}^{(0)}(\theta,\theta')$ is the value of {\bf q} when both momenta ${\bf k}$
and ${\bf k}'$
are at the Fermi surface, while ${\bf q}^{(1)}$ is the correction due to
non--zero energies $\xi$ and $\xi'$. Using standard scaling arguments we can skip 
${\bf q}^{(1)}$ in the limit $\Lambda/\epsilon_F\rightarrow 0$ because 
the cutoff is imposed to momenta. Similar argument
cannot be used for $\omega _{pp}$ and $\omega _{ph}$ because the integral in
(\ref{DlS}) runs over all frequencies independently of the actual cutoff
$\Lambda (l)$. We are therefore left with the locally dispersionless
phononic propagators
\begin{equation} \label{Dpp}
{\cal D}_l^{pp}(\theta,\theta',Q_{pp})\approx 
{\cal D}_l^{pp}(\theta,\theta', i\omega+i\omega')=
2F_l^{pp}
(\theta,\theta',i\omega+i\omega')-\tilde{F}_l^{pp}(\theta,\theta',i\omega+i\omega')
\end{equation}
and
\begin{equation} \label{Dph}
{\cal D}_l^{ph}(\theta,\theta',Q_{ph})\approx 
{\cal D}_l^{ph}(\theta,\theta', i\omega-i\omega')=
2F_l^{ph}
(\theta,\theta',i\omega-i\omega')-\tilde{F}_l^{ph}(\theta,\theta',i\omega-i\omega')\; .
\end{equation}
In these expressions we made the same pp-ph decomposition of the forward and
backward amplitudes as we did with ${\cal D}$ in eq.\ref{Decomp}. The
following step is to
re-constitute the $i\omega$ dependence from the cutoff dependence. This can
be done with logarithmic precision simply replacing 
$\Lambda$ with $i\omega$.
The derivatives over frequency in eq.(\ref{DlZ}) (acting only on $F$-parts) are then
readily calculated 
\begin{equation}\label{Diom}
\partial _{i\omega}F_l^{pp,ph}
(\theta,\theta',i\omega\pm i\omega')|_{i\omega=0}=\pm \frac{1}{i\omega'}\partial _l F_l^{pp,ph}
(\theta,\theta')\; ,
\end{equation}
and equivalently for backward amplitudes $\tilde{F}_l^{pp,ph}$. The
frequency--independent
quantities $\partial _lF_l^{pp}(\theta,\theta')$ and 
$\partial _lF_l^{ph}(\theta,\theta')$ are
 the p-p and p-h parts of the  $\beta$--function of the N--patch model,
with appropriate configurations of the external momenta:
\begin{equation} 
\label{SveBete}
\begin{array}{llll}
{\partial _lF_l^{pp}(\theta,\theta')=\beta _{pp}\{ U,U\} 
(\theta,\theta',\theta)} \\
{\partial _l\tilde{F}_l^{pp}(\theta,\theta')=\beta _{pp}\{ U,U\} 
(\theta,\theta',\theta')} \\
\partial _lF_l^{ph}(\theta,\theta')=-[ X\beta _{ph}\{ XU,XU\} ]
(\theta,\theta',\theta) \\
\partial _l\tilde{F}_l^{ph}(\theta,\theta')=[ 2\beta _{ph}\{ U,U\}
-\beta _{ph}\{ U,XU\}- \beta _{ph}\{ XU,U\}](\theta,\theta',\theta')\; ,
\end{array}
\end{equation}
where all $\beta$--functions are given in ref.\cite{ZS_prb_00}, but for the
moment with  dressed Green functions (\ref{GF_Z}). Notice that the forward
scattering has finite p-h contributions only from the ZS' channel (1=3) while
only the ZS channel (1=4) contributes to the backward scattering. This means
that we forget about the contributions at zero momentum transfer. They are 
somewhat tricky, but don't have any logarithmic part so that we can forget
them.

We can also get rid of the $Z$ factors in beta--functions of the 
eq.(\ref{SveBete})  by rescaling
the fermions at every step of the RG in a way that 
\begin{equation} \label{resc_psi}
\bar{\Psi}(\theta)Z_l^{-1}(\theta)\Psi(\theta)\rightarrow \bar{\Psi}(\theta)\Psi(\theta)
\end{equation}
and re--defining the effective interaction
\begin{equation} \label{resc_U}
U_l(1,2,3)\rightarrow[Z_l(1)Z_l(2)Z_l(3)Z_l(4)]^{-1/2}U_l(1,2,3)\; .
\end{equation}
After transformations (\ref{resc_psi}) and (\ref{resc_psi}) 
the calculations of the $\beta$--functions to the one--loop
order are identical to
the case with $Z=1$. Performing
$\omega '$  integral to
logarithmic precision, the flow equation for $Z(\theta)$ becomes
\begin{equation} \label{RG_polelog}
\partial _l\log{Z}_l(\theta)=\frac{1}{(2\pi)^2}\int d\theta'\; {\cal J}_-
(\theta',\Lambda)\eta_l(\theta,\theta')\equiv \eta_l(\theta)\; ,
\end{equation}
where
$$
\eta_l(\theta,\theta')\equiv \partial_l\{ 2[F_l^{pp}-F_l^{ph}]-
\tilde{F}_l^{pp}+\tilde{F}_l^{ph} \} (\theta,\theta')
$$
with $\partial_l F$--terms given by eqs.(\ref{SveBete}) and calculated with
the {\em bare} Green functions, as in ref.\cite{ZS_prb_00}.
Generalization of eq.\ref{RG_polelog} to finite temperatures can be done
simply by replacing ${\cal J}_-$ with $\sum _{\nu} (-\nu) {\cal
J}_{\nu}n_F(\nu\Lambda)$. 

Taking 1D ``limit''
of the eq.(\ref{RG_polelog}) is simple and instructive: instead of N patches we now have 2 patches:
$\theta=R$ (right) and $\theta=L$ (left). Independent scattering amplitudes
at non-rational filling
are $F(RL)=g_2=U(RLR)$ and $\tilde{F}(RL)=g_1=u(RLL)$. We skip $g_4=u(RRR)$
from considerations because it has no logarithmic renormalization. 
The $\theta$--integrals reduce to summation over two points so that it is 
easy to reproduce the well--known result
$\eta=-\frac{1}{4\pi^2 v_F^2}(g_1^2-g_1g_2+g_2^2)$
\cite{Q1D_Bourbonnais}. It is the Luttinger liquid exponent.

In two
dimensions the angle resolved $\eta _l(\theta)$ can also be associated with some
non-Landau (non-Fermi liquid) behavior. 
We see that $\eta $ becomes finite if the forward and
backward amplitudes have some logarithmic flow over a wide range of
$\theta$--space $(\Delta \theta \sim 1)$. Another possibility for having
finite $\eta$ is when the Fermi surface is close to the van Hove singularities. 
``Close'' means that the  distance between the Fermi level and the
van Hove singularity is comparable or inferior to the scale $\Lambda _c$ at which
interactions start to flow strongly.

We will now calculate the renormalization of $Z_l(\theta)$ in the 2D Hubbard
model at half--filling,
from the knowledge of the scale dependence of the patch-dependent 
interaction $U_l(1,2,3)$.
In the RG  equations of the 
previous section the discretization of $\theta$ is done in a way described in
ref.\cite{ZS_prb_00}. In the present case the Fermi surface is square so
that there are two mechanisms for the suppression of quasiparticle
residues. Namely, both above mentioned conditions are fulfilled:
(i) forward and backward amplitudes have logarithmic flows for any
configuration $(\theta,\theta')$ if the two angles are at opposite sides of the
Fermi surface, so that the available  phase space is indeed large;
(ii) van Hove singularities are at the Fermi surface and are {\em nested}. 
In fact,
one can alternatively imagine a Fermi surface with non-nested van Hove
singularities and nested parts elsewhere. Such a model would be even closer to
the realistic situation in some HTC compounds.
For the sake of rigor, we will however remain limited to the  Hubbard model.

\begin{figure}[htbp]
  \begin{center}
    \setlength{\unitlength}{1cm}
    \begin{picture}(7,8)
        \put(-0.5,0){\includegraphics[width=8cm]{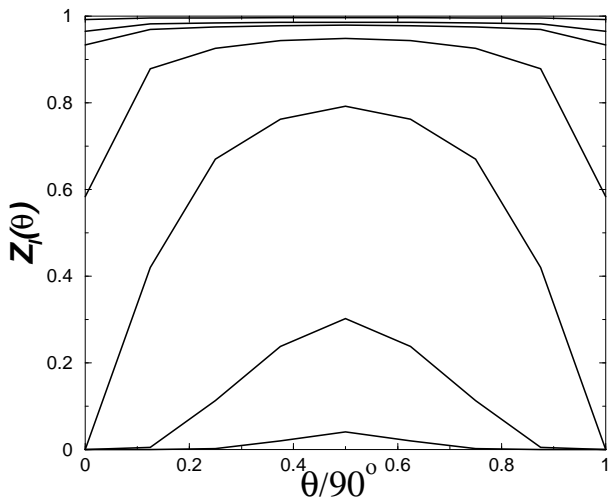}}
    \end{picture}
    \caption{The evolution of the angle--resolved quasiparticle weight on the Fermi
surface. The lines are for $l\equiv\log{(4t/\Lambda)} =$3.; 4.; 4.4; 4.95;
5.11; 5.17; 5.20. The critical scale is $l_c\approx 5.204$}
    \label{slika_z}
  \end{center}
\end{figure}

The result  is shown in fig.\ref{slika_z}. 
The figure shows $Z(\theta_i)$ with $0\leq\theta_i\leq\pi/2$
on 9 equidistant points. Different lines
correspond to different values of the scaling parameter as it approaches its
critical point, i.e. when couplings diverge. 
Settings are the same as in ref.\cite{ZS_prb_00}:
the Fermi surface was discretized
into 32 patches and the initial interaction
is $U_0/(4t)=0.333$.  One sees that the Fermi surface is first destroyed
at the van Hove points and than the regions of the FS destroyed by correlations
grow larger and larger. This kind of flow is compatible with the interpretation
that $\Lambda_c$ is not the critical temperature for some symmetry braking, but
merely the scale at which coherent quasiparticle cease to exist at the Fermi surface
giving place
to a gapped or pseudogapped liquid. 
The magnitude of the (pseudo-)gap is
largest in van Hove points and smallest on the diagonals of the Brillouin zone.
The whole gap function $\Delta(\theta)$ then
scales as $\Lambda_c\times
f(\theta)$, where $f(\theta)$ is a function with the symmetry
of the absolute value of the $d_{x^2-y^2}$--harmonic.
This  is the angle--resolved (pseudo-)gap,
responsible for the correlation--induced angle--resolved localization. In
other words the
electrons near the van Hove points get much less mobile than those near
diagonals.
The similar scenario has been
proposed by the Z\"{u}rich group \cite{Honerkamp_01}
even without concrete calculations of the quasiparticle weight.

\begin{figure}[htbp]
  \begin{center}
    \setlength{\unitlength}{1cm}
    \begin{picture}(7,7)
        \put(-0.5,0){\includegraphics[width=8cm]{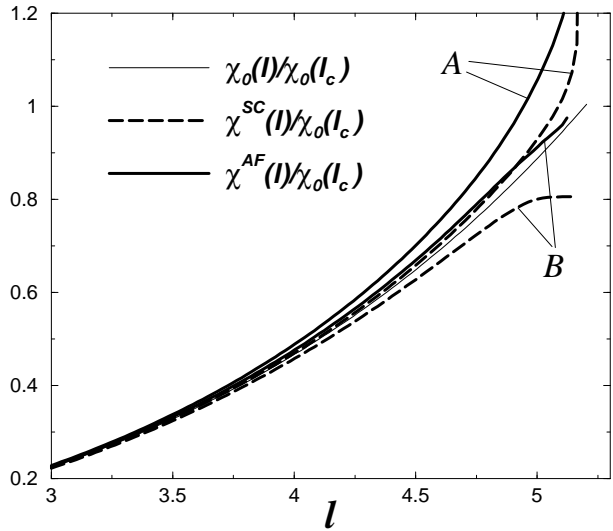}}
    \end{picture}
    \caption{Scale dependence of the dominant components of 
both antiferromagnetic and superconducting susceptibilities at  half--filling,
A: due to renormalized vertex only and  B: due to renormalized
    vertex and quasiparticle weight. The thin line represents the 
    bare susceptibility.}
    \label{susc}
  \end{center}
\end{figure}

The
question of antiferromagnetism and superconductivity remains to be
clarified. Let's discuss this problem taking into account the scale dependent
$Z(\theta)$ in the flow equations for  the susceptibilities  $\chi
^{AF}(\theta,\theta')$ and  $\chi
^{SC}(\theta,\theta')$.
Following the procedure given in ref.\cite{ZS_prb_00} and dressing the
electronic propagators with
$Z$--factors we get 
\begin{equation} \label{flow_chi}
\dot{\chi}_l^{\delta}(\theta_1,\theta_2)=\frac{1}{Z_l(\theta_1)Z_l(\theta_2)}
\oint d\theta\; 
\tilde{z}_{l}^{\delta}(\theta_1,\theta ) D_l^{\delta}(\theta ) 
\tilde{z}_{l}^{\delta}(\theta
,\theta_2) \; .
\end{equation}
This equation has the same structure as the one in ref.\cite{ZS_prb_00},
with two modifications. First, we skip the retardation effects, replacing
$l_{\delta}$ simply by $l$, because we are at  half--filling. Second, the quantity
$\tilde{z}_{l}^{\delta}(\theta_1,\theta )$ that has the role of a triangular vertex
is somewhat modified. Its flow writes:
\begin{equation} \label{flow_z}
\left[ \partial _l-\eta(\theta_1)-\eta(\theta_2)\right] 
\tilde{z}_l^{\delta}(\theta_1,\theta_2)=-\oint d\theta \; 
\tilde{z}_{l}^{\delta}(\theta_1,\theta ) D_l^{\delta}(\theta )
V_{l}^{\delta}(\theta ,\theta_2) \; .
\end{equation}
The meaning of $\tilde{z}_l^{\delta}(\theta_1,\theta_2)$ is that
$$
\tilde{z}_l^{\delta}(\theta_1,\theta_2)\equiv Z_l(\theta_1)
{z}_l^{\delta}(\theta_1,\theta_2)Z_l(\theta_2)
$$
so that the initial conditions for $\tilde{z}_l^{\delta}$ and for
${z}_l^{\delta}$ are the same. After  discretization we integrate numerically
equations (\ref{flow_chi}) and (\ref{flow_z}).
Fig.\ref{susc}
 shows the flow of the dominant eigenvalues of susceptibilities 
$\delta=AF$ and $\delta=SC$ near the divergence of scattering amplitudes.
The thin line represents both (degenerated) susceptibilities for $U=0$.
Including only the one--loop vertex renormalization we get  the strong
enhancement and, as far as my numerics can say, 
even divergences of both AF and SC susceptibilities. If we
now include also the one--particle--weight renormalization, 
both susceptibilities are radically
reduced and lose their divergent behavior. 
On the other hand, the flow of the compressibility $\chi _c$ and
magnetic susceptibility $\chi _{\sigma}$ is not affected by
$Z$--renormalization because of the Ward identities.\cite{Metzner_Ward}
All above results support the statement that what happens at energy scale
$\Lambda _c$ is a flow towards a state with spin-- and charge--gap or
pseudogap, insulating and without AF or SC ordering. 

We finish this discussion with a few words about the effects of
doping. Two regimes, {\em parquet} and BCS, exist also in the scaling properties of
$Z_l(\theta)$. In the whole parquet regime we expect the behavior governed by the
proximity to the half--filling situation, so that the present results can be 
applied at all energies larger than the chemical potential $|\mu|$. 
Upon doping the nesting becomes
more and more imperfect, the p-h logarithm loses its divergence, the
critical scale is more and more suppressed and the p-h and p-p channels get
progressively less coupled. Eventually at strong enough doping and low enough
$\Lambda$ one is  in the BCS regime
where the effective physics is described by the 2D BCS theory, with
renormalized and $\theta$--dependent interaction and quasiparticle weight. 
The anomalous dimension $\eta(\theta)$ goes to zero because 
(i) the
 the range $\Delta\theta$,
over which the forward and the backward amplitude have strong flow 
due to BCS diagram, scales with the cutoff, and (ii) the van Hove
singularities are outside of the cutoff.
The
critical temperature for the onset of the superconductivity is {\em not}
affected by the renormalized $Z$, but the magnitude and angular dependence of
the superconducting order parameter {\em are} dressed by $Z(\theta)$. The
symmetry of the gap remains $d_{x^2-y^2}$.

To summarize, I proposed a simple renormalization group theory for the angle--dependent
destruction of the Fermi surface in the Hubbard model. The results offer a
theoretical comprehension of the angle--dependent Fermi surface truncation
 in the cuprate superconductors in terms of the scattering processes of the
electrons on the low--energy collective excitations of {\em both}
particle--particle and particle--hole types.
The theory, based on the N--patch model, 
is in its essence a controlled
weak--coupling procedure that keeps trace of the dependence of
the effective interaction and one--particle spectral weight on the position
of the particles at the Fermi surface. 
As one approaches the critical scale $\Lambda _c$ the quasiparticle weight goes to zero first
near the van Hove points, and  the effect progresses toward Brillouin
zone diagonals as one lowers the temperature. 
Dressing the flow equations for AF and SC response functions with 
the one--particle weight factors
results in dramatical reduction of correlations of both types. 
The strongly correlated state just below 
$\Lambda_c$ is gapped or pseudogapped and without any long--range--order. 
Critical scale 
$\Lambda_c$ is interpreted as the pseudogap temperature $T^*$ found in 
cuprate superconductors.

\begin{acknowledgments}
I am grateful to 
Serguei Brazovskii, Benoit Dou\c{c}ot, 
Benedikt Binz, Claude Bourbonnais,  
Nicolas Dupuis, and Sebastien Dusuel for important  discussions and comments. 
Laboratoire de Physique Th\'eorique et Hautes Energies,
Universit\'es Paris VI Pierre et Marie Currie -- Paris VII Denis Diderot,
is supported by CNRS as Unit\'e Mixte de Recherche, UMR7589.
\end{acknowledgments}


\end{document}